\documentclass[twocolumn,prd,showpacs,preprintnumbers,amsmath,amssymb]{revtex4}

\usepackage{dcolumn}
\usepackage{mathrsfs}
\usepackage{bm}
\usepackage{amsmath,amssymb,epsfig,float}
\newcommand{\proj}[2]{\left|| {#1} \right\rangle\rangle\left\langle\langle {#2} \right||}

\begin{document}

\title{Quantum discord and measurement-induced disturbance in the background of dilaton black holes}
\author{Jieci Wang$^{1,2}$\footnote{jieciwang@iphy.ac.cn}, Jiliang Jing$^{1}$ and Heng Fan$^{2}$\footnote{ hfan@iphy.ac.cn}}
\affiliation{$^1$Department of Physics and Key Laboratory of Low
Dimensional \\Quantum Structures and Quantum
Control of Ministry of Education,\\
 Hunan Normal University, Changsha, Hunan 410081, People's Republic of China\\
  $^2$Beijing National Laboratory for Condensed Matter Physics,
  Institute of Physics,Chinese Academy of Sciences, Beijing 100190,
  People's Republic of China}

\vspace*{0.2cm}
\begin{abstract}

\vspace*{0.2cm}

We study the dynamics of classical correlation, quantum discord and measurement-induced disturbance of Dirac fields in the background of a dilaton black hole. We present an alternative  physical interpretation of
single mode approximation for Dirac fields in black hole spacetimes. We show that the classical and
quantum correlations are degraded as the increase of black hole's dilaton. We find that, comparing to the inertial systems, the quantum correlation measured by the one-side measuring discord is always not symmetric with respect to the measured subsystems, while the measurement-induced disturbance is symmetric. The symmetry of classical correlation and quantum discord is influenced by gravitation produced by the dilaton of the black hole.

\end{abstract}

\vspace*{1.5cm}
 \pacs{03.67.-a,03.67.Mn, 04.70.Dy}

\keywords{Quantum discord, Measurement-induced disturbance, Dirac fields, Hawking effect.}

\maketitle

\section{introduction}

Relativistic quantum information \cite{Peres}, which is
the combination of general relativity, quantum
field theory and quantum information, has been a
focus of research for both conceptual and experimental reasons.
Understanding quantum effects in a relativistic framework
is ultimately necessary because the world is essentially noninertial.
Also,  quantum correlation plays a prominent role in the study of the
thermodynamics and  information loss problem \cite{Bombelli-Callen,Hawking-Terashima} of black holes.
It is of a great interest to study  how the relativistic
effects influence the properties of entanglement, classical correlation,  quantum
discord, quantum nonlocality, as well as Fisher information \cite{Schuller-Mann, Alsing-Milburn, jieci3,jieci4, Ralph,RQI6,
adesso2,RQI3,RQI4,RQI5,RQI7} in the last few years. Recently, exotic
classes of black holes derived from the string
theory, i.e., the dilaton black holes \cite{Horowitz,Anabalon,Amarilla}
 formed by gravitational systems coupled to Maxwell and dilaton fields, have
 attracted much attention. It is widely believed that the study of dilaton
 black holes may lead to a deeper understanding of quantum gravity because
 it emerges from several fundamental theories, such as string theory, black
 hole physics, and loop quantum gravity.

On the other hand, quantum correlation measured by
discord \cite{zurek, Maziero2,
datta-onequbit,experimental-onequbit} is regarded as a
valuable resource for quantum computation and
communication in some situations. To calculate the discord of a
bipartite state, one makes a one-side measurement on a subsystem $A$ of
$\rho_{AB}$ by a complete
set of projectors $\{{\Pi^{A}_j}\}$ which yields $\rho_{B|j}
={Tr_A(\Pi^{A}_j\rho_{AB}\Pi^{A}_j)}/{p_j}$ with $
p_j=Tr_{AB}(\Pi^{A}_j\rho_{AB}\Pi^{A}_j)$. The mutual information
\cite{RAM} of $\rho_{AB}$ can alternatively be defined by ${\cal
J}_{\{\Pi^{A}_j\}}(A:B)= S(\rho_B) -S_{\{\Pi^{A}_j\}}(B|A), $ where
$S_{\{\Pi^{A}_j\}}(B|A) = \sum_j p_j S(\rho_{B|j})$  is conditional
entropy \cite{Cerf} of the state. This quantity strongly depends on
the choice of the measurements $\{\Pi^{A}_j\}$. One should minimize
the conditional entropy over all possible measurements on $A$ which
corresponds to finding the measurement that disturbs least the
overall quantum state \cite{zurek}. The quantum discord between parts
$A$ and $B$ has the form
${\cal D}(A:B)={\cal I}(A:B)-{\cal C}(A:B)$,
where ${\cal C}(A:B)$ is the classical correlation
${\cal C}(A:B)=\max_{\{\Pi^{A}_j\}}{\cal J}_{\{\Pi^{A}_j\}}(A:B)$ and
${\cal I}(A:B)$ is the quantum mutual information quantifying
the total correlation. So quantum discord describes
the discrepancy between total correlation and classical correlation,
and it thus provides a measure of quantumness of correlations.
In most situations of inertial systems, the quantum discord is
symmetric with respect to the subsystem to be measured
\cite{zurek, Maziero2,Horodecki2}. However, is the symmetry still
tenable in the noninertial systems, especially in the curved spacetimes?
Besides, does the spacetime background also affects the quantum correlations by some other measures such as the measurement-induced disturbance (MID)?

In this paper we discuss the properties of classical correlation,
quantum discord and MID \cite{Maziero3} for free modes in the background
 of a dilaton
black hole \cite{Horowitz}. The study of relativistic quantum
information on accelerated  free modes has its own advantages other than that of local
modes \cite{RQI5,RQI6} in the understanding of quantum effects
in curved spacetimes in the sense that there are no proved feasible localized
detector models inside the event horizon of a black hole. 
We assume that two observers, Alice and Bob, measure their
local state respectively. After sharing an entangled initial state,
 Alice stays stationary at an asymptotically flat region,
 while Bob moves with uniform acceleration and hovers near the
 event horizon of the dilaton black hole. We calculate the classical correlation and quantum discord by making one-side measurements on a subsystem of the bipartite system, and then get the MID measurement correlations by measuring both of the two subsystems. We are interested in how the dilaton charge will
influence the classical correlation, quantum discord, and MID, as well as if these
correlations are symmetric under the effect of gravitation produced
by the dilaton of the black hole.

The paper is organized as follows.  In the next section
we discuss the quantization of Dirac fields in the background of the dilaton black hole beyond single mode approximation \cite{Bruschi, Bruschi1}.
In Sec. III we study the properties of classical correlation, quantum discord, and MID for Dirac fields in the dilaton spacetime. We will summarize and discuss our
conclusions in the last section.

\section{Quantization of Dirac fields in dilaton black hole spacetimes}

The massless Dirac equation in a general background spacetime can be written
 as \cite{Brill}
\begin{equation}\label{Di}
[\gamma^a e_a{}^\mu(\partial_\mu+\Gamma_\mu)]\Psi=0,
\end{equation}
where $\gamma^a$ are the Dirac matrices, the four-vectors $e_a{}^\mu$
represent the inverse of the tetrad $e^a{}_\mu$ defined by
$g_{\mu\nu}=\eta_{ab}e^{a}{}_{\mu}e^b{}_{\nu}$ with $\eta_{ab}={\rm
diag}(-1, 1, 1, 1)$, $\Gamma_\mu=
\frac{1}{8}[\gamma^a,\gamma^b]e_a{}^\nu e_{b\nu;\mu}$ are the spin
connection coefficients.

The metric for a Garfinkle-Horowitz-Strominger dilaton black hole spacetime can be expressed
as
 \cite{Horowitz}
 \begin{eqnarray}
 ds^2=&-&\left(\frac{r-2M}{r-2\alpha}\right)dt^2+\left(\frac{r-2M}{r-2\alpha}\right)^{-1}
 dr^2\nonumber\\&+&r(r-2\alpha)d
 \Omega^2,\label{gem1}
 \end{eqnarray}
where $M$ and $\alpha$ are the mass of the black hole
and the dilaton. Throughout this paper we set
$G=c=\hbar=\kappa_{B}=1$. This black hole has two singular points at $r=2M$ and $r=2\alpha$. Besides, the dilaton  $\alpha$ and the mass $M$ of the black hole should satisfy $\alpha<M$. In order to separate the Dirac equation,
we  adopt a tetrad as
\begin{eqnarray}
e^a{}_\mu=diag\bigg(\sqrt{f},\frac{1}{\sqrt{f}},\sqrt{r\tilde{r}},\sqrt{r\tilde{r}}\sin\theta\bigg),
\end{eqnarray}
where $f=(r-2M)/\tilde{r}$ and $\tilde{r}=r-2\alpha$. Then Eq. (\ref{Di}) in the Garfinkle-Horowitz-Strominger dilaton black-hole spacetime becomes
\begin{eqnarray}
&-&\frac{\gamma_0}{\sqrt{f}}\frac{\partial \Psi}{\partial
t}+\sqrt{f} \gamma_1 \left[\frac{\partial }{\partial
r}+\frac{r-\alpha}{r\tilde{r}}+\frac{1}{4 f} \frac{d f}{d r} \right]
\Psi\\&+&\frac{\gamma_2}{\sqrt{r\tilde{r}}}(\frac{\partial } {\partial
\theta}+\frac{1}{2}\cot\theta)\Psi\nonumber\\&+&\frac{\gamma_3}{\sqrt{r\tilde{r}}
\sin\theta}\frac{\partial \Psi}{\partial \varphi}=0. \label{Di1}
\nonumber\\
\end{eqnarray}
If we rescale $\Psi$ as $\Psi=f^{-\frac{1}{4}}\Phi$ and use an ansatz
for the Dirac spinor similar to Ref. \cite{jieci1}, we can solving
the Dirac equation near the event horizon. For the outside and
inside region of the event horizon, we obtain the positive
frequency outgoing solutions \cite{D-R,jieci2}
\begin{eqnarray}\label{outside mode}
\Psi^{+}_{out,\mathbf{k}}=\mathcal {G}e^{-i\omega \mathcal {U}},
\end{eqnarray}
\begin{eqnarray}\label{inside mode}
\Psi^{+}_{in,\mathbf{k}}=\mathcal {G}e^{i\omega \mathcal {U}},
\end{eqnarray}
where $\mathcal {U}=t-r_{*}$ and  $\mathcal{G}$ is a four-component
Dirac spinor, $\mathbf{k}$ is the wavevector we used to label
the modes hereafter and for massless Dirac field
$\omega=|\mathbf{k}|$.
 In terms of these basis the Dirac field $\Psi$ can be expanded as
\begin{eqnarray}\label{First expand}
\Psi&=&\int
d\mathbf{k}[\hat{a}^{out}_{\mathbf{k}}\Psi^{+}_{out,\mathbf{k}}+\hat{b}^{out\dag}_{\mathbf{-k}}
\Psi^{-}_{out,\mathbf{k}}\nonumber\\ &+&\hat{a}^{in}_{\mathbf{k}}\Psi^{+}_{in,\mathbf{k}}
+\hat{b}^{in\dag}_{\mathbf{-k}}\Psi^{-}_{in,\mathbf{k}}],
\end{eqnarray}
where $\hat{a}^{out}_{\mathbf{k}}$ and $\hat{b}^{out\dag}_{\mathbf{k}}$
are the fermion annihilation and antifermion creation operators
acting on the state of the exterior region, and
$\hat{a}^{in}_{\mathbf{k}}$ and $\hat{b}^{in\dag}_{\mathbf{k}}$ are
the fermion annihilation and antifermion creation operators acting
on the interior vacuum of the black hole respectively.
The annihilation operator $\hat{a}^{out}_{\mathbf{k}}$ and  creation
operator $\hat{a}^{out}_{\mathbf{k}}$ satisfy the canonical anticommutation
$\{\hat{a}^{out}_{\mathbf{k}},\hat{a}^{out}_{\mathbf{k'}}\}=\delta_{\mathbf{k}\mathbf{k'}}$ and  $\{\hat{a}^{out}_{\mathbf{k}},\hat{a}^{out\dagger}_{\mathbf{k'}}\}
=\{\hat{a}^{out\dagger}_{\mathbf{k}},\hat{a}^{out\dagger}_{\mathbf{k'}}\}=0$, where $\{.,.\}$ denotes the anticommutator. Clearly,  two
fermionic Fock are antisymmetric with respect to the exchange
of the mode labels $\mathbf{k}$ and $\mathbf{k'}$ due to the anticommutation relations. We therefore define
\begin{eqnarray}||1_{\mathbf{k}}1_{\mathbf{k'}}\rangle\rangle=
\hat{a}^{\dagger}_{\mathbf{k}},\hat{b}^{\dagger}_{\mathbf{k'}}||0\rangle\rangle
=-\hat{b}^{\dagger}_{\mathbf{k'}},\hat{a}^{\dagger}_{\mathbf{k}}||0\rangle\rangle,
\end{eqnarray}
where the states in the antisymmetric fermionic Fock space are denoted  by double-lined Dirac notation $||.\rangle\rangle$ \cite{antifermi} rather than  the single-lined notations.

Making analytic continuation for Eqs. (\ref{outside mode}) and
(\ref{inside mode}), we find a complete basis for positive energy
modes, i.e., the Kruskal modes, according to the suggestion of Domour-Ruffini \cite{D-R}. Then we can quantize the massless Dirac field in black hole and Kruskal modes respectively \cite{jieci1,jieci2}, from which we can easily get the Bogoliubov transformations \cite{Barnett} between the creation and annihilation operators in different
coordinates \cite{jieci1}. Considering that it is more interesting to quantize the
Dirac field beyond the single mode approximation \cite{Bruschi, Bruschi1}. We
construct a different set of creation operators that are linear
combinations of creation operators in the inside and outside regions \cite{Bruschi, Bruschi1}
\begin{eqnarray}\label{Dirac-ex}
\nonumber\tilde{c}_{\mathbf{k},R} &=& \cos r \hat{a}^{out}_{\mathbf{k}} - \sin r
\hat{b}^{in\dagger}_{\mathbf{-k}}, \nonumber\\
 \tilde{c}_{\mathbf{k},L} &=& \cos r \hat{a}^{in}_{\mathbf{k}}  - \sin r
\hat{b}^{out\dagger}_{\mathbf{-k}},\nonumber\\
\nonumber\tilde{c}^\dagger_{\mathbf{k},R} &=& \cos r \hat{a}^{out\dagger}_{\mathbf{k}} - \sin r
\hat{b}^{in}_{\mathbf{-k}}, \nonumber\\
 \tilde{c}^\dagger_{\mathbf{k},L} &=& \cos r \hat{a}^{in\dagger}_{\mathbf{k}}  - \sin r
\hat{b}^{out}_{\mathbf{-k}}, \label{umodes}
\end{eqnarray}
where $\cos r = [e^{-8\pi\omega
(M-\alpha)}+1]^{-\frac{1}{2}}$ and $\sin r  = [e^{8\pi\omega
(M-\alpha)}+1]^{-\frac{1}{2}}$.  We name the modes (or operators) with subscripts L and R by ``left" and  ``right"  modes (or operators), respectively. After properly normalizing the state vector, the
Kruskal vacuum is found to be $|| 0 \rangle\rangle_K = \bigotimes_{\mathbf{k}} || 0_{\mathbf{k}} \rangle\rangle_{K}=\bigotimes_{\mathbf{k}}||0_{\mathbf{k}} \rangle\rangle_{R}\otimes||0_{\mathbf{k}} \rangle\rangle_{L}$, where $|||0_{\mathbf{k}} \rangle\rangle_{R}$ and $|||0_{\mathbf{k}} \rangle\rangle_{L}$ are annihilated by the annihilation operators $\tilde{c}_{\mathbf{k},R}$ and $\tilde{c}_{\mathbf{k},L}$. The vacuum state $||0_{\mathbf{k}}\rangle\rangle_{K}$ for mode $\mathbf{k}$ is given by
\begin{eqnarray}\label{Dirac-vacuum}
||0_{\mathbf{k}}\rangle\rangle_{K}&=&\cos^2 r||0000\rangle\rangle
-\sin r ||0011\rangle\rangle\nonumber\\&+&\sin r\cos r||1100\rangle\rangle-\sin^2 r||1111\rangle\rangle,
\end{eqnarray}
where $||mnm'n'\rangle\rangle=||m_{\mathbf{k}}\rangle\rangle^{+}_{out}
||n_{-\mathbf{k}}\rangle\rangle^{-}_{in}
||m'_{-\mathbf{k}}\rangle\rangle^{-}_{out}
||n'_{\mathbf{k}}\rangle\rangle^{+}_{in}$, $\{||n_{-\mathbf{k}}\rangle\rangle^{-}_{in}\}$ and
$\{||n_{\mathbf{k}}\rangle\rangle^{+}_{out}\}$ are the orthonormal bases for
the inside and outside region of the dilaton
black hole respectively, and the $\{+,-\}$ superscript on the kets
is used to indicate the fermion and antifermion vacua.
For the observer Bob who travels outside the event horizon, the Hawking radiation spectrum from the view of  his
detector can be obtained by
\begin{eqnarray}\label{Hawking}
\nonumber N_\omega^{2}=_{K}\langle0|\hat{a}^{out\dag}_{\mathbf{k}}
\hat{a}^{out}_{\mathbf{k}}|0\rangle_{K}=\frac{1}{e^{\omega/T }+1},
\end{eqnarray}
where $T=\frac{1}{8\pi(M-\alpha)}$ is Hawking temperature \cite{Hawking} of the black hole.
This equation shows that the observer in the exterior of the
Garfinkle-Horowitz-Strominger dilaton black hole detects a thermal Fermi-Dirac distribution
of particles.
Because of the Pauli exclusion principle, only the first excited state for each
fermion mode $||1_{\mathbf{k}}\rangle\rangle^{+}_{K}$ is allowed,
and similarly for antifermions.
The first excited state for the fermion mode is given by
\begin{eqnarray}\label{Dirac-excited}
||1_{\mathbf{k}}\rangle\rangle^{+}_{K}&&=
\left[q_R (\tilde{c}^\dagger_{\mathbf{k},R}\otimes I_L) + q_L (I_R\otimes \tilde{c}^\dagger_{\mathbf{k},L}) \right]||0_{\mathbf{k}}\rangle\rangle_{K}\nonumber\\
&&=q_R[\cos r||1000\rangle\rangle
-\sin r ||1011\rangle\rangle]\nonumber\\
&&+q_L[\sin r||1100\rangle\rangle+\cos r||0001\rangle\rangle],
\end{eqnarray}
with $ |q_R|^2 + |q_L|^2 = 1$. The study of fermionic
quantum information beyond the single mode approximation, which was
proposed in \cite{Bruschi} and widely adopted recently, has a lack of physical interpretation so far. Here we present an alternative physical
interpretation on the operators and states that obtained beyond such an approximation in black hole spacetimes. The operator $\tilde{c}^\dagger_{\mathbf{k},R}$ in Eq.(\ref{Dirac-ex})
indicates the creation of two particles, i.e., a fermion in the
exterior vacuum and an antifermion in the interior vacuum of the
black hole. Similarly, the create operator $\tilde{c}^\dagger_{\mathbf{k},L}$
means that an antifermion and an fermion are created outside and
inside the event horizon, respectively.  Hawking radiation comes from spontaneous creation of paired particles and
antiparticles by quantum fluctuations near the event horizon.
The particles and antiparticles can
radiate toward the inside and outside regions randomly from the event
horizon with the total  probability $ |q_R|^2 + |q_L|^2 = 1$.
Thus, $|q_R|=1$  means that all the particles move
toward the black hole exteriors while all the antiparticles move
to the inside region, i.e., only particles can be detected as
Hawking radiation. Similarly, $|q_L|=1$ indicates that only antiparticles
escapes from the event horizon.
Therefore, the single mode approximation (either $|q_R|=1$ or $|q_L|=1$) is a special case when either only particles or only antiparticles are detected.

\section{Quantum discord and MID in dilaton black hole spacetimes}

 We assume that Alice and Bob share a maximally entangled state
\begin{eqnarray}\label{initial}
||\Phi\rangle\rangle_{AB}=\frac{1}{\sqrt{2}}(||0\rangle\rangle_{A}||0\rangle\rangle_{B}
+||1\rangle\rangle^+_{A}||1\rangle\rangle^+_{B}),
\end{eqnarray} at the same point in the
asymptotically flat region of the dilaton black hole.
Then Alice stays stationary at the asymptotically flat region,
while Bob moves with uniform acceleration and hovers near the
event horizon of the dilaton black hole.
 Bob  will detects a thermal Fermi-Dirac distribution of particles
 and his detector is found to be excited. Using Eqs.
(\ref{Dirac-vacuum}) and (\ref{Dirac-excited}), we can rewrite Eq.
(\ref{initial}). Since Bob is causally
disconnected from the region inside the event horizon we should
trace over the state of the inside region and obtain
\begin{eqnarray}
\nonumber\varrho^{AB_{out}}&=&\frac{1}{2}\Big[C^4\proj{000}{000}+S^2C^2(\proj{010}{010}\\
&+&\proj{001}{001})\nonumber+\phantom{\Big[}S^4\proj{011}{011}\\
\nonumber&+&|q_\text{\text{R}}|^2(C^2\proj{110}{110}+S^2\proj{111}{111})\\*
\nonumber&+&\phantom{\Big[}|q_\text{\text{L}}|^2(S^2\!\proj{110}{110}\!+\!C^2\!\proj{100}{100})
\nonumber\\&+&q_\text{\text{R}}^*(C^3\!\proj{000}{110}
\nonumber+S^2C\proj{001}{111})\\ &-&\nonumber q_\text{\text{L}}^*(C^2S\!\proj{001}{100}+S^3\!\proj{011}{110})\\*
&&\!\!-q_\text{R} q_\text{L}^*SC\!\proj{111}{100}\Big]+(\text{H.c.})_{_{\substack{\text{non-}\text{diag.}}}}
\end{eqnarray}
where $||lmn\rangle\rangle=||l\rangle\rangle_{A}
||m_{\mathbf{k}}\rangle\rangle^{+}_{out}
||n_{\mathbf{-k}}\rangle\rangle^{-}_{out}$, $S=\sin r$ and
$C=\cos r$.  We assume that Bob has a
detector sensitive only to the particle modes, which means
that an antifermion cannot be excited in a single detector when
a fermion was detected. Thus, we should also trace out the antifermion
mode $||n_{\mathbf{-k}}\rangle\rangle^{-}_{out}$
in the outside region \cite{Bruschi1}
\begin{eqnarray}  \label{eq:state1}
\varrho^{A\tilde{B}}&=&\frac{1}{2}\bigg[C^2||
00\rangle\rangle\langle\langle00|+
q^*_\text{R}C||00\rangle\rangle\langle\langle11||\nonumber\\
&+&q_\text{R}C||11\rangle\rangle\langle\langle00||)
+q_\text{L}C^2||10\rangle\rangle\langle\langle10|| \nonumber\\
&+&S^2
||01\rangle\rangle\langle\langle01||+\chi_0||11\rangle\rangle\langle\langle11||\bigg],
\end{eqnarray}
where $||lm\rangle\rangle=||l\rangle\rangle_{A}
||m_{\mathbf{k}}\rangle\rangle^{+}_{out}$
and $\chi_0=|q_\text{R}|^2+|q_\text{L}|^2 S^2$. Hereafter we call the mode
$\||m_{\mathbf{k}}\rangle\rangle^{+}_{out}$ as  $\tilde{B}$. Now our system
has two subsystems, i.e., the inertial subsystem $A$ and accelerated
subsystem $\tilde{B}$. We can easily obtain the
von Neumann entropy $S(\rho_{A,\tilde{B}})$ of this state, $S(\rho_A)$ for
the reduced density matrix of the mode $A$
and $S(\rho_{\tilde{B}})$ for the mode  $\tilde{B}$, respectively.

\subsection{Measurements on subsystem $A$}

Now let us first make measurements on the subsystem $A$, the projectors
are defined as \cite{zurek,datta-onequbit,jieci3}
\begin{eqnarray}\label{statem1}
\Pi^{A}_{+}=\frac{I_2+\mathbf{n}\cdot\mathbf{\sigma}}{2}\otimes I_2,\ \ \ \ \
\Pi^{A}_{-}=\frac{I_2-\mathbf{n}\cdot\mathbf{\sigma}}{2}\otimes I_2,
\end{eqnarray}
where $n_1=\sin\theta\cos\varphi,\ n_2=\sin\theta\sin\varphi,\
n_3=\cos\theta$ and $\sigma_i$ are Pauli matrices. The measurement operators in Eq. (\ref{statem1}) include a two-outcome projective measurement operator $\frac{I_2\pm\mathbf{n}\cdot\mathbf{\sigma}}{2}$ on subsystem $A$  and an identity operator on subsystem $B$. These operators are orthogonal projectors spanning the qubit Hilbert space and can therefore be parameterized by the
unit vector $\mathbf{n}=(n_1,n_2, n_3)$. For simplicity, here we can take the measurements on the particle-number degree of freedom, i.e., to measure whether or not a fermion with wave vector $\mathbf{k}$ is excited in the particle detector. After the
measurement of $\Pi^{A}_{+}$, the quantum state $\rho_{A,\tilde{B}}$ changes to
\begin{eqnarray}\label{stama1}
\nonumber\rho^{MA}_{+}&=&Tr_{A}(\Pi^{A}_{+}\rho_{A,\tilde{B}}\Pi^{A}_{+})/p^{A}_{+}\\
&=&\frac{1}{2}\left[
     \begin{array}{cc}
       \varsigma_1 & e^{i\varphi}q^*_\text{R}C \sin\theta  \\
       e^{-i\varphi} q_\text{R} C\sin\theta  & \chi_1\\
     \end{array}
   \right],
\end{eqnarray}
where $p^{A}_{+}=Tr(\Pi^{A}_{+}\rho_{A,\tilde{B}} \Pi^{A}_{+})=1/2$,
$\varsigma_1=C^{2}[1+\cos\theta+q_\text{L}(1-\cos\theta)]$
and $\chi_1=\chi_0(1-\cos\theta)+(1+\cos\theta)S^2$.
The same method is used to compute the state after measurement $\Pi^{A}_{-}$, then we have
\begin{eqnarray}\label{stama2}
\rho^{MA}_{-}=\frac{1}{4p^{A}_{-}}\left[
     \begin{array}{cc}
       \varsigma_2 & -e^{i\varphi}q^*_\text{R}C \sin\theta  \\
       -e^{-i\varphi} q_\text{R} C\sin\theta  & \chi_2\\
     \end{array}
   \right],
\end{eqnarray}
where $p^{A}_{-}=Tr(\Pi^{A}_{-}\rho_{A,\tilde{B}} \Pi^{A}_{-})=1/2$,
$\varsigma_2=C^{2}[1-\cos\theta+q_\text{L}(1+\cos\theta)]$
and $\chi_2=\chi_0(1+\cos\theta)+(1-\cos\theta)S^2$.
Now we can obtain the conditional entropy $S_{\{\Pi^{A}_j\}}(\tilde{B}|A)\equiv\sum_jp_j
S(\tilde{B}|j)$. The classical correlation in this case is
\begin{eqnarray}\label{classical1}
\nonumber{\cal C}(\tilde{B}|A)=S(\rho_{\tilde{B}})
-\min_{\Pi^{A}_j}S_{\{\Pi^{A}_j\}}(\tilde{B}|A),
\end{eqnarray}
and the quantum discord is
\begin{eqnarray}\label{discord1}
\nonumber {\cal D}(\tilde{B}|A)=S(\rho_{\tilde{A}})-S(\rho_{A,\tilde{B}})+\min_{\Pi^{A}_j}S_{\{\Pi^{A}_j\}}(\tilde{B}|A).
\end{eqnarray}

Note that the conditional entropy has to be numerically evaluated by optimizing over the angles $\theta$ and $\phi$. Thus we should
minimize it over all possible measurements on $A$ \cite{zurek}. We find that the condition entropy is independent of
$\varphi$ and its minimum can be obtained when $\theta=\pi/2$.

\subsection{Measurements on subsystem ${\tilde{B}}$}

Then let us make our measurements on the  subsystem ${\tilde{B}}$; the projectors
are defined as
\begin{eqnarray}
\Pi^{\tilde{B}}_{+}=I_2\otimes\frac{I_2+\mathbf{n}\cdot\sigma}{2} ,\ \ \ \ \
\Pi^{\tilde{B}}_{-}=I_2\otimes\frac{I_2-\mathbf{n}\cdot\sigma}{2}.
\end{eqnarray}
After the measurement of $\Pi^{\tilde{B}}_{+}$, the state $\rho_{A,\tilde{B}}$ changes to
\begin{eqnarray}\label{con1}
\nonumber\rho^{M\tilde{B}}_{+}
=\frac{1}{4p^{\tilde{B}}_{+}}\left[
     \begin{array}{cc}
      \varsigma_3&~~~~ e^{i\varphi}q^*_\text{R} \sin r\sin\theta  \\
       e^{-i\varphi}q_\text{R} \sin r \sin\theta  &  \chi_3 \\
     \end{array}
   \right],
\end{eqnarray}
where $p^{\tilde{B}}_{+}
=(1 -\cos 2r \cos\theta)(1+
q^2_\text{L})+q^2_\text{R}(1 + \cos\theta)$, $\varsigma_3= (1+\cos\theta)C^{2}+
(1-\cos\theta)S^2$ and $\chi_3=q^2_\text{L}C^2(1-\cos\theta)
+\chi_0(1+\cos\theta)$. Similarly, we can calculate  the state after $\Pi^{\tilde{B}}_{-}$
and get the classical correlation ${\cal C}(A|\tilde{B})$
and the quantum discord ${\cal D}(A|\tilde{B})$ respectively.

\begin{figure}[ht]
\includegraphics[scale=0.6]{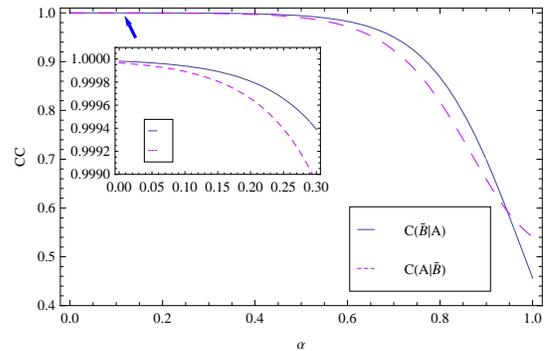}
\caption{\label{ClaT}(Color online) The classical
correlation ${\cal C}(\tilde{B}|A)$ obtained by measuring the subsystem $A$ and
the classical correlation ${\cal C}(A|\tilde{B})$ (dashed  line)
as a function of the dilaton$\alpha$. We set $M=\omega=1$ and $q_R=1$.}
\end{figure}

Figure \ref{ClaT} shows how the dilaton $\alpha$ of the black hole influences
the classical correlations of the system
when we obtain them by measuring the subsystem $A$ and ${\tilde{B}}$,
respectively. From which we can see that for all the two cases the classical
correlations decrease with increasing dilaton $\alpha$.  The classical
correlation ${\cal C}(\tilde{B}|A)$ obtained by the one-side measurements
on subsystem $A$, is \textit{always not equal} to ${\cal C}(A|\tilde{B})$
for any dilaton value. In contrast, the classical
correlations satisfy ${\cal C}(B|A)={\cal C}(A|B)=1$ for
the initial state, Eq.(\ref{initial}), in the asymptotically
flat region. Comparing  to the flat spacetime, the classical
correlation (of course the related quantum correlation) is \textit{not
symmetrical} in the dilaton black hole spacetime. This asymmetry is due
to the effect of gravitation produced by the black hole. We also note that ${\cal C}(\tilde{B}|A)$
is larger than ${\cal C}(A|\tilde{B})$ when the dilaton is smaller
than a fixed value ($\alpha\simeq 0.9451$), while it is
smaller than ${\cal C}(A|\tilde{B})$ when the dilaton is
larger than this value.

\subsection{Symmetric measurement of the correlations}

From the foregoing discussion, we see that the classical and quantum
correlations in the curved spacetime depend on the measurement process.
At the same time, a symmetric measurement of the quantum correlation was proposed recently. The MID
measurement \cite{Maziero3}, which is obtained  by a complete set of
projective measurements over both partitions of a
bipartite state,
is given by ~\cite{Maziero3}
\begin{equation} \label{sQC}
{\cal D}(MID)=\mathcal{I}(\rho_{A,\tilde{B}}) - \mathcal{I}(\eta_{A,\tilde{B}})
\end{equation}
with
\begin{equation}
 \label{E:midstate}
\eta_{A,\tilde{B}}=\sum_{i=1}^m\sum_{j=1}^n\mathbf{\pi}_{i}^{A}\otimes\mathbf{\pi}_{j}^{\tilde{B}}
\rho_{A,\tilde{B}}\mathbf{\pi}_{i}^{A}\otimes\mathbf{\pi}_{j}^{\tilde{B}}.
\end{equation}
where  $\mathbf{\pi}_{i}^{A}=||i\rangle\rangle\langle\langle i||$ and
$\mathbf{\pi}_{j}^{\tilde{B}} =||j\rangle\rangle\langle\langle j||$ are
one-dimensional orthogonal projections for parties $A$ and
$\tilde{B}$, respectively.  Such a symmetrized version
of the quantum correlation was recently discussed theoretically
 ~\cite{Maziero3} and experimentally measured by anuclear magnetic
resonance setup at room temperature \cite{NMR}. Besides, MID requires only the local measurement strategy rather than the cumbersome optimization required by the derivation of discord \cite{Rao}.
Here we can easily obtain
$\eta_{A,\tilde{B}}$ and find the MID measure of
classical correlation ${\cal C}(MID)$ and quantum
correlation ${\cal D}(MID)$.

\begin{figure}[ht]
\includegraphics[scale=0.6]{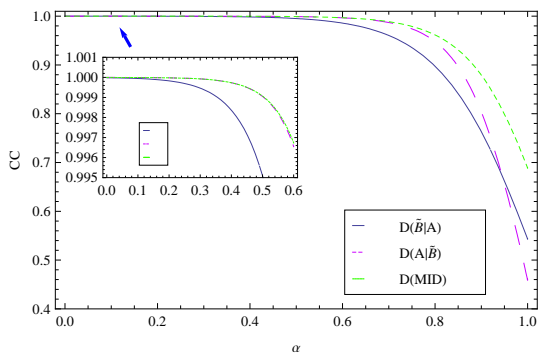}
\caption{\label{DisT}(Color online) The quantum discord ${\cal D}(\tilde{B}|A)$(solid line),
${\cal D}(A|\tilde{B})$ (dashed line),
and ${\cal D}(MID)$ (dotted line) of $\rho_{A,\tilde{B}}$
as a function of the dilaton $\alpha$. We set $M=\omega=1$ and $q_R=1$.}
\end{figure}

Figure \ref{DisT} shows how the dilaton of the black hole affects
the quantum correlations (discord and MID) that obtained by different
measuring methods. Both the quantum discord and the MID decrease as the increasing of
 $\alpha$, which means the quantum correlations degraded as the increase
of dilaton $\alpha$. It is shown that the
discord ${\cal D}(A|\tilde{B})$ is not equal to ${\cal
D}(\tilde{B}|A)$ for any $\alpha$, which is extremely different from ${\cal D}(B|A)={\cal D}(A|B)=1$ in the initial state Eq. (\ref{initial}). Comparing to the flat spacetime, the quantum discord is not symmetric for any $\alpha$ in the dilaton black hole spacetime. The symmetry of quantum discord is truly
influenced by the dilaton of the black hole. 
It is also noted that the quantum correlation obtained via the MID measurement
is always larger than that obtained by the one-side measurement.

\section{summary}

The effect of black hole's dilaton on the symmetry of
classical correlation, quantum discord, and MID for the Dirac
fields is investigated. We give a physical interpretation of
the single mode approximation in the curved spacetime, i.e.,
such an approximation is a special case when either only particles
or only antiparticles are detected. It is shown that
the classical and quantum correlations decrease
monotonously as increasing dilaton, which means all type of correlations are
degraded due to the effect of gravitation generated by the dilaton of the
black hole. We find  that both the one-side measured classical and quantum
correlations are not symmetric with respect to the subsystem being measured.
So both the quantification and
the symmetry of classical correlation and quantum discord are influenced by the
gravitation when taking the one-side measurement.
This is a sharp comparison between the inertial systems
and the system in the curved spacetime.
The results obtained here are not only
helpful to understanding the symmetric properties of classical
and quantum correlations in the presence of strong gravitation but also give a
better insight into quantum properties of dilaton black holes.

\begin{acknowledgments}
This work is supported by the 973 program through
2010CB922904, the National Natural Science Foundation
of China under Grant No. 11305058, No.11175248, No.11175065,  the Doctoral Scientific Fund Project of the Ministry of Education of China under Grant No. 20134306120003, and CAS.
\end{acknowledgments}

\end{document}